\documentclass[10pt]{article}
\usepackage[dvips]{graphicx}
\begin{document}
\begin{center}
\textbf{{\large Coupled Quintessence and Phantom Based On Dilaton}}
 \vskip 0.35 in
\begin{minipage}{4.5 in}
\begin{center}
{\small Z. G. HUANG \vskip 0.06 in \textit{
Department~of~Mathematics~and~Physics,
\\~Huaihai~Institute~of~Technology,~222005,~Lianyungang,~China
\\
zghuang@hhit.edu.cn}} \vskip 0.25 in {\small H. Q. LU and W. FANG \vskip 0.06
in \textit{
Department~of~Physics,~Shanghai~University,~Shanghai,~China
\\
alberthq$\_$lu@staff.shu.edu.cn}}
\end{center}
\vskip 0.2 in

{\small Based on dilatonic dark energy model, we consider two cases:
dilaton field with positive kinetic energy(coupled quintessence) and
with negative kinetic energy(phantom). In the two cases, we
investigate the existence of attractor solutions which correspond to
an equation of state parameter $\omega=-1$ and a cosmic density
parameter $\Omega_\sigma=1$. We find that the coupled term between
matter and dilaton can't affect the existence of attractor
solutions. In the Mexican hat potential, the attractor behaviors,
the evolution of state parameter $\omega$ and cosmic density
parameter $\Omega$, are shown mathematically. Finally, we show the
effect of coupling term on the evolution of
$X(\frac{\sigma}{\sigma_0})$ and
$Y(\frac{\dot{\sigma}}{\sigma^2_0})$ with respect to $N(lna)$
numerically. \vskip 0.2 in \textit{Keywords:} Dark energy; Dilaton;
Quintessence; Attractor solution; Phantom
\\
}
\end{minipage}
\end{center}
\vskip 0.2 in
\begin{flushleft}\textbf{1. Introduction}\end{flushleft}
Observational data from SNe Ia[1], WMAP[2], SDSS[3],
show that we are living in a spatially flat universe which consists of about
two thirds dark energy with negative pressure, one third dust matter
including cold dark matters plus baryons, and negligible radiation,
and that our universe is undergoing an accelerated expansion. In order to explain
current acceleration of the universe, we require an exotic energy dubbed "dark energy"
with equation of state satisfying $\omega<-\frac{1}{3}$.
\par Many models have been proposed so far to fit these observations.
Among these models, the most typical ones are cosmological constant
and a time varying scalar field with positive or negative kinetic
energy evolving in a specific potential, referred to as
"quintessence"[4-14] or "phantom"[15-19]. The essential
characteristics of these dark energy models are contained in the
parameter of its equation of state, $p=\omega\rho$, where $p$ and
$\rho$ denote the pressure and energy density of dark energy,
respectively, and $\omega$ is a state parameter. Quintessence model
has been widely studied, and its state parameter $\omega$ which is
time-dependent, is greater than $-1$. At the same time, observations
also suggest another dark energy model---phantom with an equation of
state $\omega<-1$. The most typical characteristic of the phantom
model is that the kinetic of the scalar field is negative. Because
of this, it leads to some unusual features of phantom model. A
striking consequence of phantom model is that the Universe will
undergo a catastrophic "Big Rip" in a finite time. However, Hao and
Li[20] have presented that a de Sitter attractor will prevent the
phantom energy from increasing up to infinite in a finite cosmic
time, therefor the presence of phantom energy dose not lead to a
cosmic doomsday in a theory with de Sitter attractor at late time.
On the other hand, it has been shown that big rip singularity can be
generically avoided if one considers quantum gravitational
corrections to dynamics[21].
\par Some authors have considered the coupled dark energy[22]. In the work of L.Amendola[23],
he investigated the cosmological consequences of the coupled
quintessence model, assuming an exponential potential. In our
previous paper[24], we have considered a dilatonic dark energy
model, based on Weyl-scaled induced gravitational theory. In that
paper, we find that when the dilaton field is not gravitational
clustered at small scales, the effect of dilaton can not change the
evolutionary law of baryon density perturbation, and the density
perturbation can grow from $z\sim10^3$ to $z\sim5$, which guarantees
the structure formation. When dilaton energy is very small compared
the matter energy, potential energy of dilaton field can be
neglected. In this case, the solution of cosmological scale $a$ has
been found[25]. In what follows, we shall consider a more general
dilatonic dark energy model that the contributions from radiation
and matter can't be neglected. In this model, we will study coupled
quintessence---dilaton field with positive kinetic energy, and
phantom---dilaton field with negative kinetic energy and find out
the sufficient conditions of existence of attractor solutions, which
limit the choice of potential of scalar field. When we take the
potential as the form $\frac{\mu}{4}(\sigma^2-\varepsilon^2)^2+W_0$
named Mexican hat potential, we investigate existence and dependence
on initial conditions for a late time attractor and obtain the
sufficient conditions for the existence of an attractor solutions.
We find that these attractor solutions correspond to an equation of
state $\omega=-1$ and a cosmic density parameter $\Omega_\sigma=1$,
which are important features for a dark energy model that can meet
the current observations. These results are shown mathematically.
Finally, we show the effect of coupling term on the evolution of $X$
and $Y$ with respect to $N$ numerically.
\par This paper is organized as follows: In section 2, we introduce Weyl-scaled induced
gravitational theory and regard dilaton as quintessence. Based on
this, we investigate the sufficient condition for existence of
attractor solution. In section 3, we consider phantom model and give
the sufficient condition for existence of attractor solution.
Section 4 is a special example---Mexican hat potential. Section 5 is
summery.
 \vskip 0.2 in
\begin{flushleft}\textbf{2. Coupled Quintessence}\end{flushleft}

To deduce the field equation of induced gravitational theory, let us
consider the action of Jordan-Brans-Dicke theory firstly
 \begin{equation}S=\int{d^4X\sqrt{-\gamma}[\phi
 \tilde{R}-\omega\gamma^{\mu\nu}\frac{\partial_\mu\phi\partial_\nu\phi}{\phi}-\Lambda(\phi)+\tilde{L}_{fluid}(\psi)}]\end{equation}
where the lagrangian density of cosmic fluid
$\tilde{L}_{fluid}(\psi)=\frac{1}{2}\gamma^{\mu\nu}\partial_\mu\psi\partial_\nu\psi-V(\psi)$,
$\gamma$ is the determinant of $\gamma_{\mu\nu}$ which is Jordan
metric, $\omega$ is the dimensionless coupling parameter, $R$ is the
contracted $R_{\mu\nu}$. The metric sign convention is(-,+,+,+). The
quantity $\Lambda(\phi)$ is a nontrivial potential of $\phi$ field.
When $\Lambda(\phi)\neq0$ the action of Eq.(1) describes the induced
gravity. The energy density of cosmic fluid
$\widetilde{\rho}=\frac{1}{2}(\frac{d\psi}{d\widetilde{t}})^2+V(\psi)$
and the pressure $
\widetilde{p}=\frac{1}{2}(\frac{d\psi}{d\widetilde{t}})^2-V(\psi).$
\par However it is often useful to write the action in terms of the conformally related Einstein metric. We introduce the dilaton field $\sigma$ and
conformal transformation as follows
\begin{equation}\phi=\frac{1}{2}e^{\alpha\sigma}\end{equation}
\begin{equation}\gamma_{\mu\nu}=e^{-\alpha\sigma}g_{\mu\nu}\end{equation}
where $\alpha^2=\frac{\kappa^2}{2\varpi+3}$ with $\varpi>3500$[26]
being an important parameter in Weyl-scaled induced gravitational
theory, $g_{\mu\nu}$ is the Pauli metric. In this paper, we work in
units($\kappa^2=8\pi G=c=1$). From the solar system tests, the
current constrain is $\alpha^2<0.001$[27]. The new constrain on the
parameter is $\alpha^2<0.0001$[28], which seems to argue against the
existence of long-range scalars. Perhaps such a pessimistic
interpretation of the limit is premature [27].
\par The action(1) becomes
Eq.(4) by performing the conformal transformation Eq.(2) and Eq.(3)
\begin{equation}S=\int{d^4X\sqrt{-g}[\frac{1}{2}R(g_{\mu\nu})-\frac{1}{2}g^{\mu\nu}\partial_\mu\sigma\partial_\nu\sigma-W(\sigma)+L_{fluid}(\psi)}]\end{equation}
where
$L_{fluid}(\psi)=\frac{1}{2}g^{\mu\nu}e^{-\alpha\sigma}\partial_\mu\psi\partial_\nu\psi-e^{-2\alpha\sigma}V(\psi)$.
The conventional Einstein gravity limit occurs as $\sigma\rightarrow
0$ for an arbitrary $\omega$ or $\omega\rightarrow\infty$ with an
arbitrary $\sigma$.
\par The nontrivial potential of the $\sigma$ field,
$W(\sigma)$ can be a metric scale form of $\Lambda(\phi)$.
Otherwise, one can start from Eq.(4), and define $W(\sigma)$ as an
arbitrary nontrivial potential. $g_{\mu\nu}$ is the pauli metric.
Damour and Cho et.al pointed out  that  the pauli metric can
represent the massless spin-two graviton in induced gravitational
theory. Cho also pointed out that in the compactification of
Kaluza-Klein theory, the physical metric must be identified as the
pauli metric because of the the wrong sign of the kinetic energy
term of the scalar field in the Jordan frame[29]. The dilaton field
appears in string theory naturally.
\par By varying the action Eq.(4), one can obtain the field equations
of Weyl-scaled induced gravitational theory.
$$R_{\mu\nu}-\frac{1}{2}g_{\mu\nu}R=-\frac{1}{3}\{[\partial_\mu\sigma\partial_\nu\sigma-\frac{1}{2}g_{\mu\nu}\partial_\rho\sigma\partial^\rho\sigma]
-g_{\mu\nu}W(\sigma)$$
\begin{equation}+e^{-\alpha\sigma}[\partial_\mu\psi\partial_\nu\psi-\frac{1}{2}g_{\mu\nu}\partial_\rho\psi\partial^\rho\psi]
-g_{\mu\nu}e^{-2\alpha\sigma}V(\psi)\}\end{equation}
\begin{equation}\Delta\sigma=\frac{dW(\sigma)}{d\sigma}-\frac{\alpha}{2}e^{-2\alpha\sigma}g^{\mu\nu}\partial_\mu\psi\partial_\nu\psi
-2\alpha e^{-2\alpha\sigma}V(\psi)\end{equation}
\begin{equation}\Delta\psi=-\alpha g_{\mu\nu}\partial_\mu\psi\partial_\nu\sigma+e^{-\alpha\sigma}\frac{dV(\psi)}{d\psi}\end{equation}
The energy-momentum tensor $T_{\mu\nu}$ of cosmic fluid is
\begin{equation}T_{\mu\nu}=(\rho+p)U_\mu U_\nu+pg_{\mu\nu}\end{equation}
where the density of energy
\begin{equation}\rho=\frac{1}{2}\dot{\psi}^2+e^{-\alpha\sigma}V(\psi)\end{equation}
the pressure
\begin{equation}p=\frac{1}{2}\dot{\psi}^2-e^{-\alpha\sigma}V(\psi)\end{equation}
$\rho$ and $p$ are related to their directly measurable counterparts
by $\rho=e^{-\alpha\sigma}\widetilde{\rho},
p=e^{-\alpha\sigma}\widetilde{p}.$
\par In FRW universe, the field
equations of Weyl-scaled induced gravitational theory can be
expressed as follows:
\begin{equation}H^2=\frac{1}{3}[\frac{1}{2}\dot{\sigma}^2+W(\sigma)+e^{-\alpha\sigma}\rho]\end{equation}
\begin{equation}\ddot{\sigma}+3H\dot{\sigma}+\frac{dW}{d\sigma}=\frac{1}{2}\alpha e^{-\alpha\sigma}(\rho-3p)\end{equation}
\begin{equation}\dot{\rho}+3H(\rho+p)=\frac{1}{2}\alpha\dot{\sigma}(\rho+3p)\end{equation}
where $H$ is Hubble parameter. From Eq.(12), we note that there a
coupling between matter and dilaton. Its effect will be discussed at
the end of this section. For radiation $\rho_r=3p_r$, we get
$\rho_r\propto \frac{e^{\alpha\sigma}}{a^4}$ from Eq.(13). For
matter $p_m=0$, we get $\rho_m\propto
\frac{e^{\frac{1}{2}\alpha\sigma}}{a^3}$ from Eq.(13). Taking these
results into Eqs.(11) and (12), we obtain
\begin{equation}H^2=H_i^2[\frac{\frac{1}{2}\dot{\sigma}^2+W(\sigma)}{\rho_{c,i}}+\Omega_{m,i}e^{-\frac{1}{2}\alpha\sigma}(\frac{a_i}{a})^3+\Omega_{r,i}(\frac{a_i}{a})^4]\end{equation}
\begin{equation}\ddot{\sigma}+3H\dot{\sigma}+\frac{dW}{d\sigma}=\frac{1}{2}\alpha\eta_ie^{-\frac{1}{2}\alpha\sigma}a^{-3}\end{equation}
where $H_i^2=\frac{\rho_{c,i}}{3}$, $\rho_{c,i}$ is the critical
energy density of the universe at initial time $t_i$. $H_i$,
$\Omega_{m,i}$, $\Omega_{r,i}$ denote the Hubble parameter, matter
energy density parameter, radiation energy density parameter at
initial time $t_i$ respectively, and
$\eta_i=\frac{\rho_{m,i}a_i^3}{e^{\frac{1}{2}\alpha\sigma_i}}$ with
$\rho_{m,i}$ being matter energy density at initial time $t_i$. We
define our starting point as the equipartition epoch, at which
$\Omega_{m,i}=\Omega_{r,i}=0.5$ and consider the initial scale
factor $a_i=1$ for convenience. So we have
\begin{equation}H^2=H_i^2[\frac{\frac{1}{2}\dot{\sigma}^2+W(\sigma)}{\rho_{c,i}}+\Omega_{m,i}e^{-\frac{1}{2}\alpha\sigma}a^{-3}+\Omega_{r,i}a^{-4}]\end{equation}
The effective density $\rho_{\sigma}$ and effective pressure
$p_{\sigma}$ can be expressed as follows
\begin{equation}\rho_{\sigma}=\frac{1}{2}\dot{\sigma}^2+W(\sigma)\end{equation}
\begin{equation}p_\sigma=\frac{1}{2}\dot{\sigma}^2-W(\sigma)\end{equation}
So, the equation of state of dilaton field is
\begin{equation}\omega_\sigma=\frac{\frac{1}{2}\dot{\sigma}^2-W(\sigma)}{\frac{1}{2}\dot{\sigma}^2+W(\sigma)}\end{equation}
In order to gain more insights into the dynamical system, we
introduce the new dimensionless variables
\begin{equation}X=\frac{\sigma}{\sigma_0},~~~Y=\frac{\dot{\sigma}}{\sigma_0^2},~~~N=ln~a\end{equation} Eq.(19)
becomes
\begin{equation}\omega_\sigma=\frac{\frac{\sigma_0^4Y^2}{2}-W(X)}{\frac{\sigma_0^4Y^2}{2}+W(X)}\end{equation}
 The field Eqs.(14)(15) could be rewritten as follows
\begin{equation}\frac{dX}{dN}=\frac{\sigma_0Y}{H}\end{equation}
\begin{equation}\frac{dY}{dN}=-3Y-\frac{W'(X)}{\sigma_0^3H}+\frac{\alpha\eta_ie^{-\frac{1}{2}\alpha\sigma_0X}e^{-3N}}{4H}\end{equation}
where the prime denotes the derivative with respect to $X$ and $H$
could be expressed as
\begin{equation}H=H_i[\frac{\frac{1}{2}\sigma_0^4Y^2+W(X)}{\rho_{c,i}}+\Omega_{m,i}e^{-\frac{1}{2}\alpha\sigma_0X}a^{-3N}+\Omega_{r,i}a^{-4N}]^{\frac{1}{2}}\end{equation}
\par According to reference[30], the dynamic system decided by Eqs.(22)(23) belongs to the class of so called autonomous systems. The coordinates of critical point is decided by
the following condition
\begin{equation}\frac{dX}{dN}=0,~~~~~~~~~\frac{dY}{dN}=0\end{equation}
\par In dilaton-dominant epoch, the matter energy density can be neglected
comparing to the dilaton energy density, that is $\frac{\alpha\eta_i
e^{-\frac{1}{2}\alpha\sigma_0X}e^{-3N}}{4H}\rightarrow0$. So
autonomous systems (22)(23) become
\begin{equation}\frac{dX}{dN}=\frac{\sigma_0Y}{H}\end{equation}
\begin{equation}\frac{dY}{dN}=-3Y-\frac{W'(X)}{\sigma_0^3H}\end{equation}
The critical point of the above autonomous system is $(X_c,0)$,
where $X_c$ is defined by $W'(X_c)=0$. Linearize the above equations
about the critical point, we have
\begin{equation}\frac{dX}{dN}=\frac{\sqrt{3}\sigma_0Y}{\sqrt{W(X_c)}}\end{equation}
\begin{equation}\frac{dY}{dN}=-3Y-\frac{\sqrt{3}W''(X_c)X}{\sigma_0^3\sqrt{W(X_c)}}\end{equation}
The eigenvalues of the system are
\begin{equation}x_{1,2}=\frac{-\xi\pm\sqrt{\xi^2-4\zeta}}{2}\end{equation}
where $\xi=3$ and $\zeta=\frac{W''(X_c)}{W(X_c)\sigma^2_0}$. For a
positive potential, if $W''(X_c)>0$, i.e. $x_1<0$ and $x_2<0$, the
critical point is a stable point which corresponds to a late time
attractor solution. So, $W(X_c)>0$, $W'(X_c)=0$ and $W''(X_c)>0$ is
the sufficient condition for existence of attractor solution in
quintessence model. In this paper we take potential $W(X)$ as the
Mexican hat potential which satisfies the above condition
$W(X_c)>0$, $W'(X_c)=0$ and $W''(X_c)>0$. Therefore, our
quintessence model can admit a late time attractor solution.
\par Based on the analysis of the above autonomous system, we can conclude that:
Because $\frac{\alpha\eta_i
e^{-\frac{1}{2}\alpha\sigma_0X}e^{-3N}}{4H}\rightarrow0$, the
coupling between matter and dilaton can't change existence of
attractor solution in Eqs.(22)(23), that is to say, the coupled term
can't change the fate of the evolution of the universe. In one word,
the existence of couple term $\frac{\alpha\eta_i
e^{-\frac{1}{2}\alpha\sigma_0X}e^{-3N}}{4H}$ between matter and
dilaton affects the evolutive process of the universe, but not the
fate of the universe.
\par In what follows, we shall show mathematically the evolutions
of the components of cosmic density $\Omega$, the evolution of the parameter
of state equation $\omega$ and the evolution of $X$, $Y$ with respect to $N$ in quintessence model. The attractor behavior is also shown in phase plane.
\vskip 0.2 in  \begin{flushleft}\textbf{3. Phantom Model}\end{flushleft}
When we regard dilaton as phantom field, its kinetic energy will be negative. So, the
phantom field equations become
\begin{equation}H_p^2=\frac{1}{3}[-\frac{1}{2}\dot{\sigma}^2+W(\sigma)+e^{-\alpha\sigma}\rho]\end{equation}
\begin{equation}\ddot{\sigma}+3H_p~\dot{\sigma}-\frac{dW}{d\sigma}=\frac{1}{2}\alpha e^{-\alpha\sigma}(\rho-3p)\end{equation}
Based on the same analysis like the quintessence model, we have,
\begin{equation}H^2=H_i^2[\frac{-\frac{1}{2}\dot{\sigma}^2+W(\sigma)}{\rho_{c,i}}+\Omega_{m,i}e^{-\frac{1}{2}\alpha\sigma}(\frac{a_i}{a})^3+\Omega_{r,i}(\frac{a_i}{a})^4]\end{equation}
\begin{equation}\ddot{\sigma}+3H\dot{\sigma}-\frac{dW}{d\sigma}=\frac{1}{2}\alpha\eta_i e^{-\frac{1}{2}\alpha\sigma}a^{-3}\end{equation}
According to the transition (20), the phantom equation fields become
\begin{equation}\frac{dX}{dN}=\frac{\sigma_0Y}{H}\end{equation}
\begin{equation}\frac{dY}{dN}=-3Y+\frac{W'(X)}{\sigma_0^3H}+\frac{\alpha\eta_i e^{-\frac{1}{2}\alpha\sigma_0X}e^{-3N}}{4H}\end{equation}
where the prime denotes the derivative with respect to $X$, and $H$
can be expressed as
\begin{equation}H=H_i[\frac{-\frac{1}{2}\sigma_0^4Y^2+W(X)}{\rho_{c,i}}+\Omega_{m,i}e^{-\frac{1}{2}\alpha\sigma_0X}a^{-3N}+\Omega_{r,i}a^{-4N}]^{\frac{1}{2}}\end{equation}
The equation of state parameter becomes
\begin{equation}\omega_\sigma=\frac{-\frac{\sigma_0^4Y^2}{2}-W(X)}{-\frac{\sigma_0^4Y^2}{2}+W(X)}\end{equation}
\par Eqs.(35)(36) and their initial conditions determine the evolution
of phantom universe and the behaviors of the late time de Sitter
attractor, which include the existence and the stability of the
solution. According to the similar analysis to the quintessence
model, we also conclude: $W(X_c)>0$, $W'(X_c)=0$ and
$W''(X_c)<0(x_1<0, x_2<0)$ is the sufficient condition for existence
of attractor solution in phantom model. In this paper we take
potential $W(X)$ as the Mexican hat potential which satisfies the
above condition $W(X_c)>0$, $W'(X_c)=0$ and $W''(X_c)<0$. Therefore,
our phantom model can admit a late time attractor solution.
\par In what follows, we shall show
mathematically the evolutions of the components of cosmic density
$\Omega$, the evolution of the parameter of state equation $\omega$
and the evolution of $X$, $Y$ with respect to $N$ in phantom model.
The attractor behavior is also shown in phase plane.

\vskip 0.2 in
\begin{flushleft}\textbf{4. Mexican Hat Potential $\frac{\mu}{4}(\sigma^2-\varepsilon^2)^2+W_0$}\end{flushleft}
\par For this type of Mexican hat potential, it has two extremum points in the range $\sigma\geq0$: a minimum at $\sigma=\varepsilon$ and a maximum at $\sigma=0$.
The non-conventional parameter $W_0$ in this potential, moves the potential up and down, which is equivalent to adding a cosmological
constant to the usual Mexican hat potential. We show the feature of Mexican hat potential mathematically in Fig.1.
\vskip 0.1 in
\begin{center}
\includegraphics[scale=0.5]{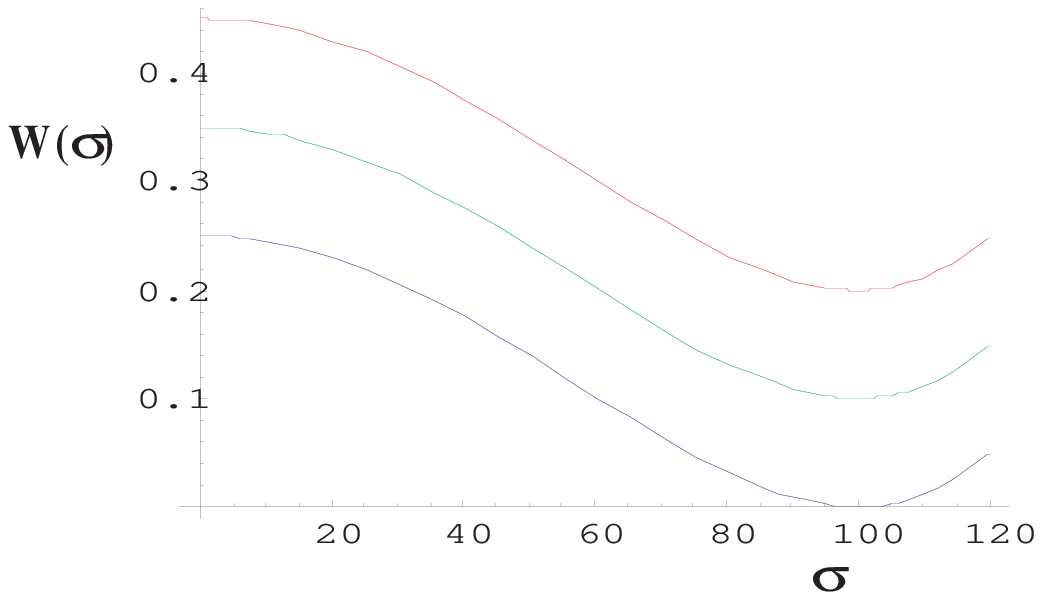}
\begin{center}{\small Fig.1 Mexican hat potential when we set $W_0=0(lower~line),~0.1(middle~line),~0.2(upper~line)$.}
\end{center}
\end{center}
\par The critical point of the above autonomous system is
$(X_c,0)$, where $X_c$ is defined by $W'(X_c)=0$. Note that the
energy density of the dilaton field at the critical point is
$W(X_c)$ and should not vanish, thus the sufficient condition for
the existence of a viable cosmological model with a late time de
Sitter attractor solution should be that: the potential of the field
has non-vanishing minimum value in quintessence model, on the
contrary, the potential should have non-vanishing maximum value in
phantom model. The extremum difference of potential between
quintessence model and phantom model results from the unusual
physical feature of phantom field.
\par Since the Mexican hat potential($W_0\neq0$) has a non-vanishing minimum value and a non-vanishing maximum value,
there must exist late time attractors in the quintessence model and
phantom model, which both drive from dilaton field. Based on this,
we study the attractor behaviors of quintessence and phantom model
in Mexican hat potential.
\par In quintessence model, the critical point is
$X_{c1}=\varepsilon$, and at this point the potential has a minimum value $W_0$. Obviously, it is an attractor solution,
which corresponds to an equation of state parameter $\omega=-1$ and a cosmic density parameter $\Omega_\sigma=1$.
The evolution of the parameter of state equation $\omega$, the cosmic density parameter $\Omega$, $Y$ with respect to $N$ are
shown numerically in Figs.2-4. We also show the attractor property for quintessence in phase plane in Fig.5.
\vskip 0.15 in
\begin{minipage}{0.4\textwidth}
\includegraphics[scale=0.6]{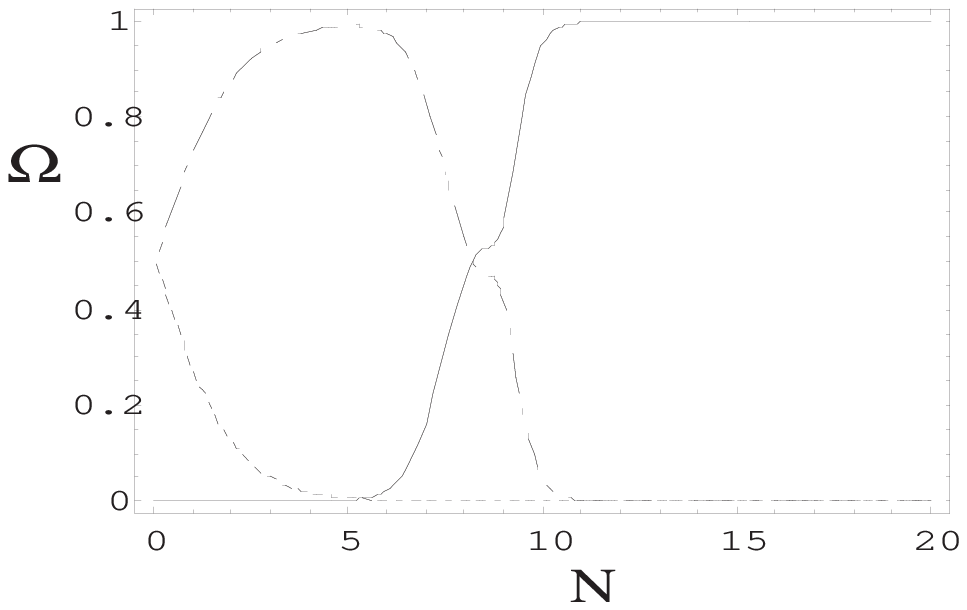}
{\small Fig.2 The evolution of cosmic density parameters
$\Omega_\sigma(solid~line)$, $\Omega_m(dash-dot~line)$,
$\Omega_r(dot~line)$ with respect to N in Mexican hat potential for quintessence. We set $\frac{\mu}{H_i}=10^{-12}$, $\alpha=0.01$, $W_0=1.00$.}
\end{minipage}
\hspace{0.1\textwidth}
\begin{minipage}{0.4\textwidth}
\includegraphics[scale=0.6]{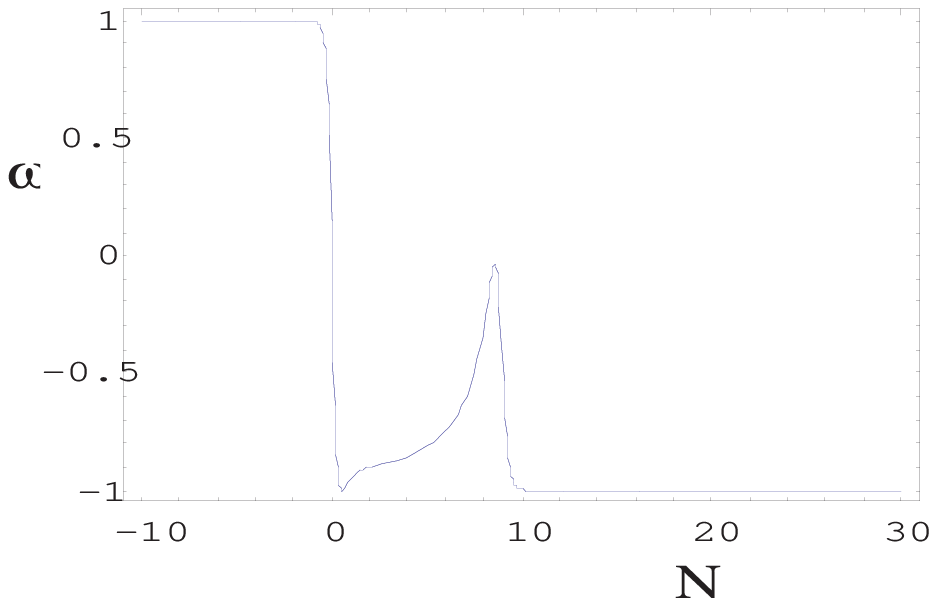}
{\small Fig.3 The evolution of the parameter of state equation
$\omega$ with respect to N in Mexican hat potential for
quintessence. We set $\frac{\mu}{H_i}=10^{-12}$, $\alpha=0.01$,
$W_0=1.00$.}
\end{minipage}

\begin{minipage}{0.4\textwidth}
\includegraphics[scale=0.6]{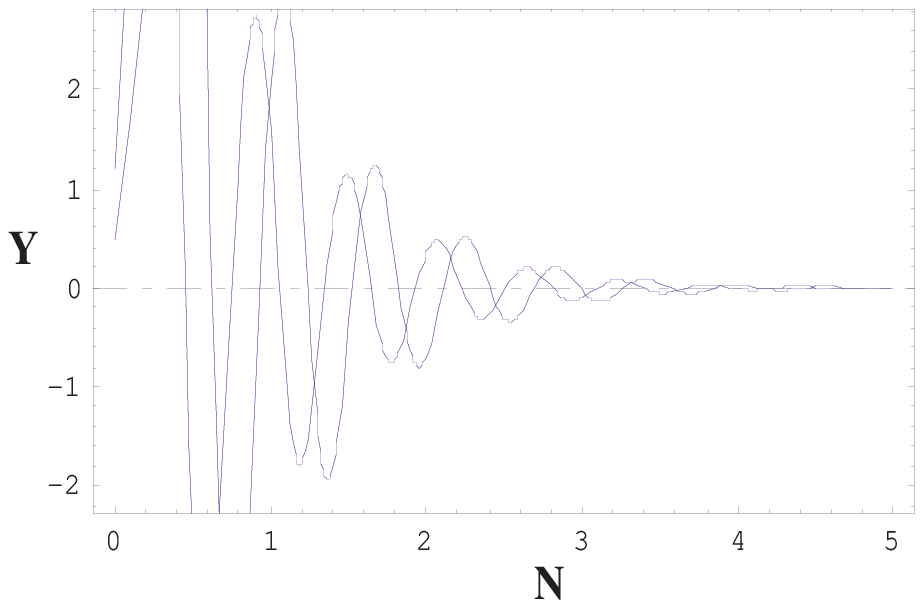}
{\small Fig.4 The evolution of Y with respect to N for different
coordinate initial conditions $X_i=0.5$, $Y_i=0.5$; $X_i=1.2$,
$Y_i=1.2$ in Mexican hat potential for quintessence.}
\end{minipage}
\hspace{0.1\textwidth}
\begin{minipage}{0.4\textwidth}
\includegraphics[scale=0.6]{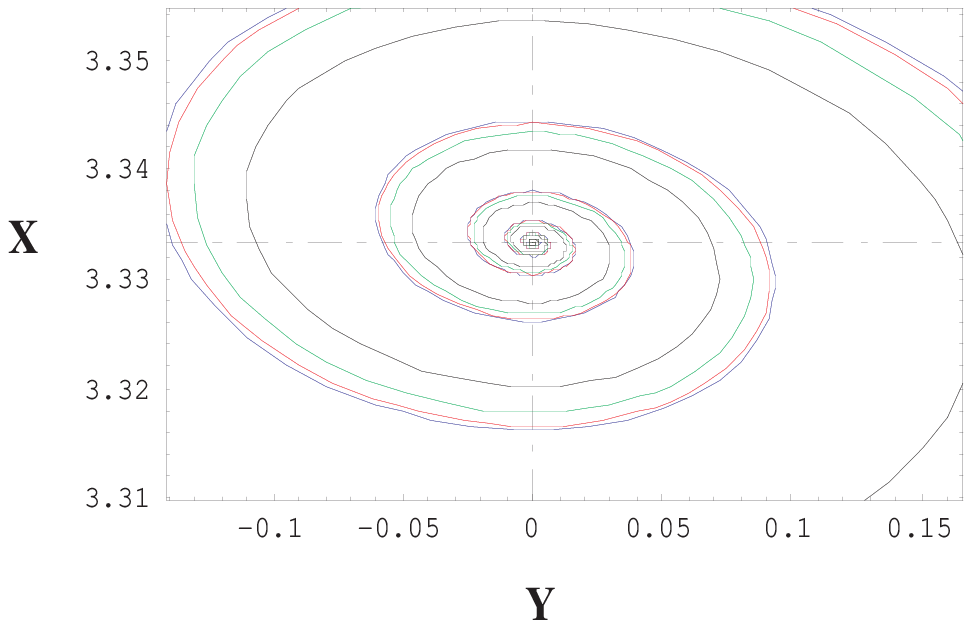}
{\small Fig.5 The phase portrait for quintessence in Mexican hat potential for several different
initial conditions. The critical point $X_c=3.333$ corresponds to a de Sitter solution which is a stable spiral}
\end{minipage}

\vskip 0.3 in
\par In phantom model, according to the sufficient condition of existence of attractor solution, we can obtain
the critical point is $X_{c2}=0$, and at this point the potential has a maximum value $\frac{\mu\varepsilon^4}{4}+W_0$.
Obviously, it is also an attractor solution, which corresponds to an equation of state parameter $\omega=-1$ and a cosmic density parameter $\Omega_\sigma=1$.
The evolution of the parameter of state equation $\omega$ and the cosmic density parameter $\Omega$ with respect to $N$ are
shown numerically in Fig.6 and Fig.7. We also show the attractor property for phantom in phase plane in Fig.8.
\vskip 0.3 in

\begin{minipage}{0.4\textwidth}
\includegraphics[scale=0.6]{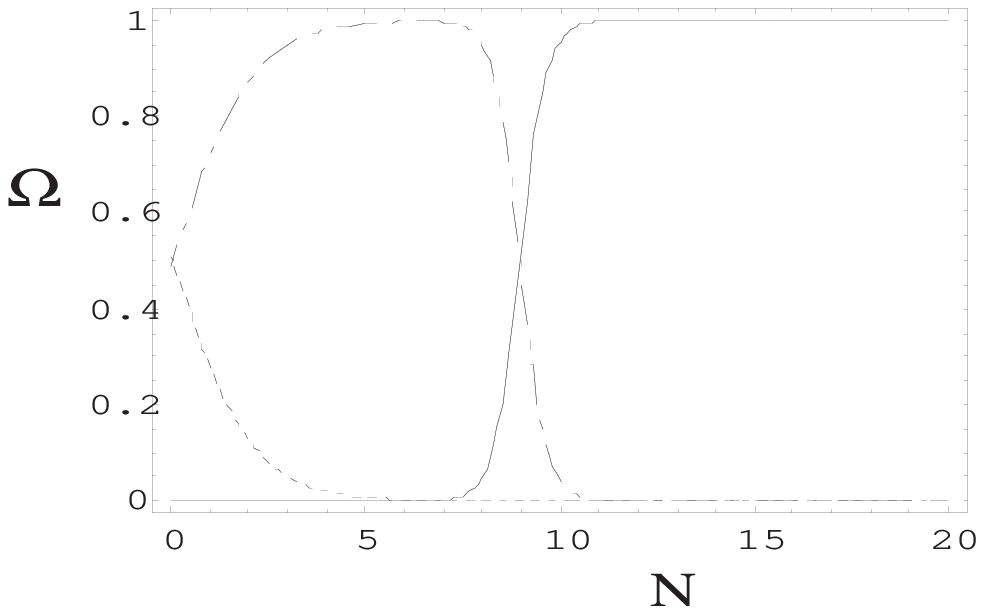}
{\small Fig.6 The evolution of cosmic density parameters
$\Omega_\sigma(solid~line)$, $\Omega_m(dash-dot~line)$,
$\Omega_r(dot~line)$ with respect to N in Mexican hat potential for phantom.
We set $\frac{\mu}{H_i}=10^{-15}$, $\alpha=0.01$, $W_0=1.00$.}
\end{minipage}
\hspace{0.1\textwidth}
\begin{minipage}{0.4\textwidth}
\includegraphics[scale=0.6]{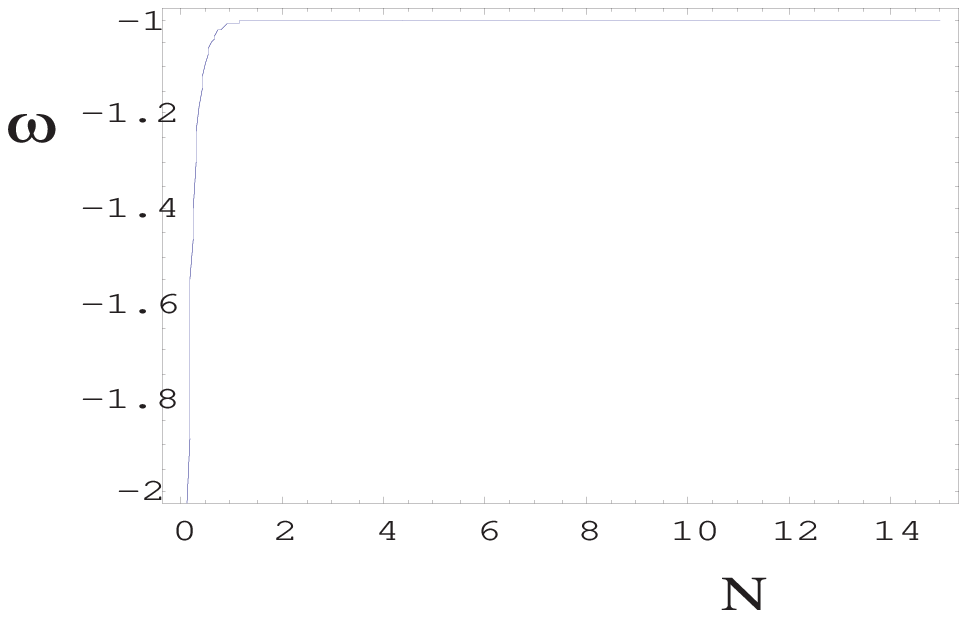}
{\small Fig.7 The evolution of the parameter of state equation
$\omega$ with respect to N in Mexican hat potential for phantom. We
set $\frac{\mu}{H_i}=10^{-15}$, $\alpha=0.01$, $W_0=1.00$.}
\end{minipage}

\begin{center}
\includegraphics[scale=0.6]{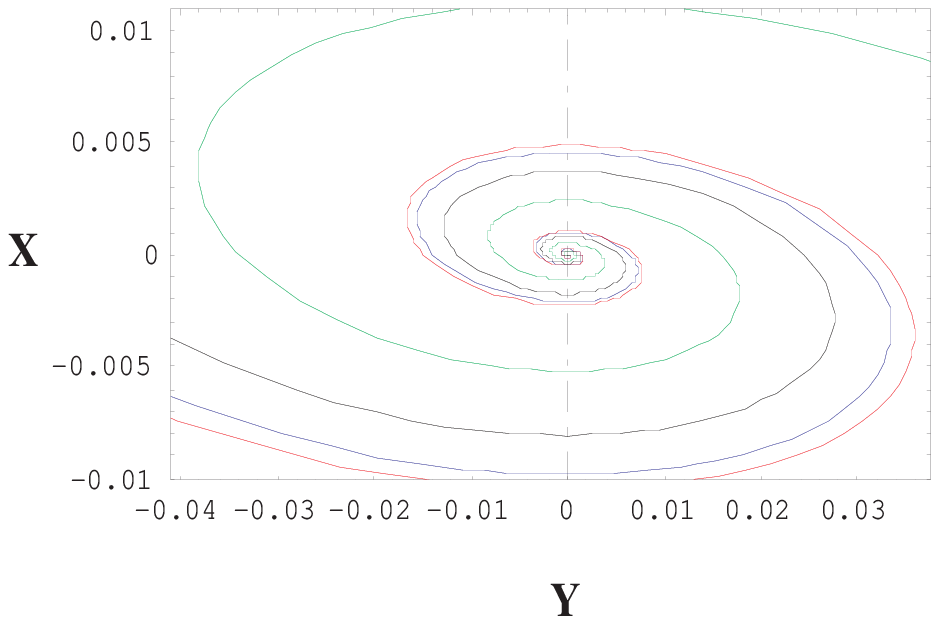}
\begin{center}{\small Fig.8 The phase portrait for phantom in Mexican hat potential for several different
initial conditions. The critical point $X_c=0$ corresponds to a de Sitter solution which is a stable spiral}
\end{center}
\end{center}
\par In what follows, we shall show the effect of nonminimum coupling term $\frac{\alpha\eta_i e^{-\frac{1}{2}\alpha\sigma_0X}e^{-3N}}{4H}$ on the
evolution of $X$, $Y$ and the attractor solution. From Figs.9-11 and the analysis in Section 2, we obtain that the existence of couple term
between matter and dilaton affects the evolution course of the universe but not the result. That is to say, either quintessence or phantom
model admits an attractor in spite of the existence of coupling term.

\begin{minipage}{0.4\textwidth}
\includegraphics[scale=0.6]{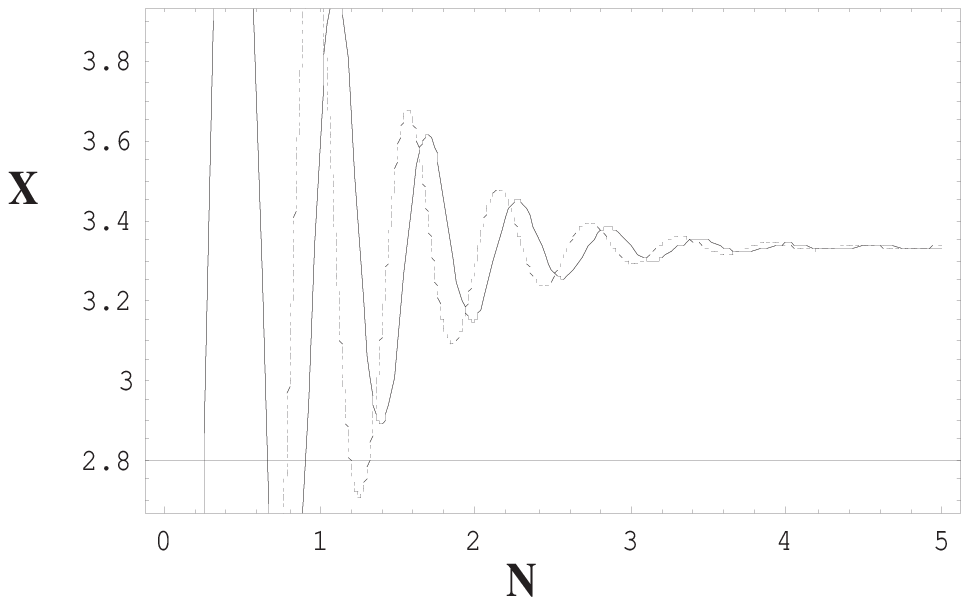}
{\small Fig.9 The evolutive comparison of $X$ with respect to $N$ in
quintessence with coupling term(real line) and without coupling
term(dot line).}
\end{minipage}
\hspace{0.1\textwidth}
\begin{minipage}{0.4\textwidth}
\includegraphics[scale=0.6]{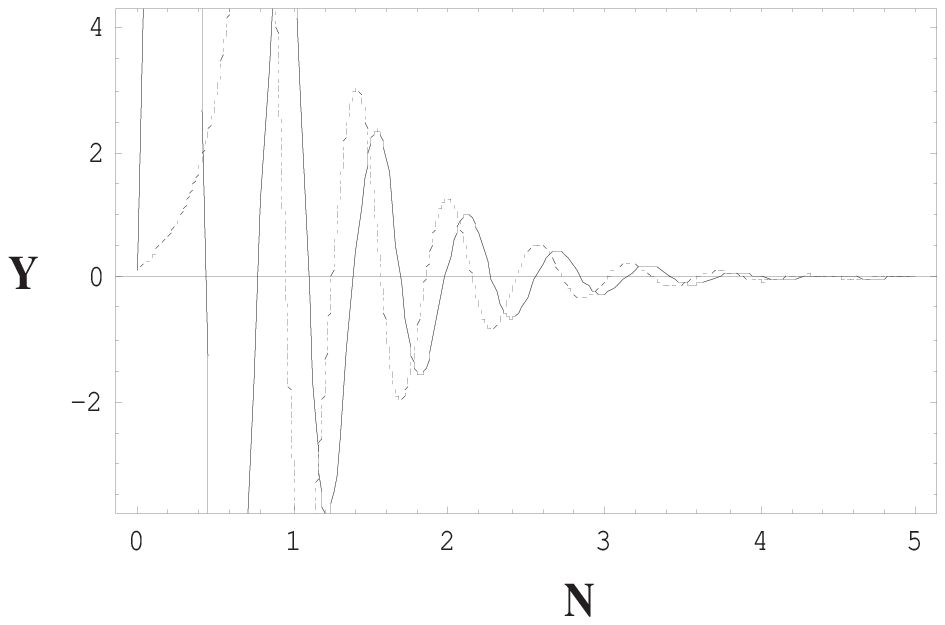}
{\small Fig.10 The evolutive comparison of $Y$ with respect to $N$
in quintessence with coupling term(real line) and without coupling
term(dot line).}
\end{minipage}
\begin{center}
\includegraphics[scale=0.6]{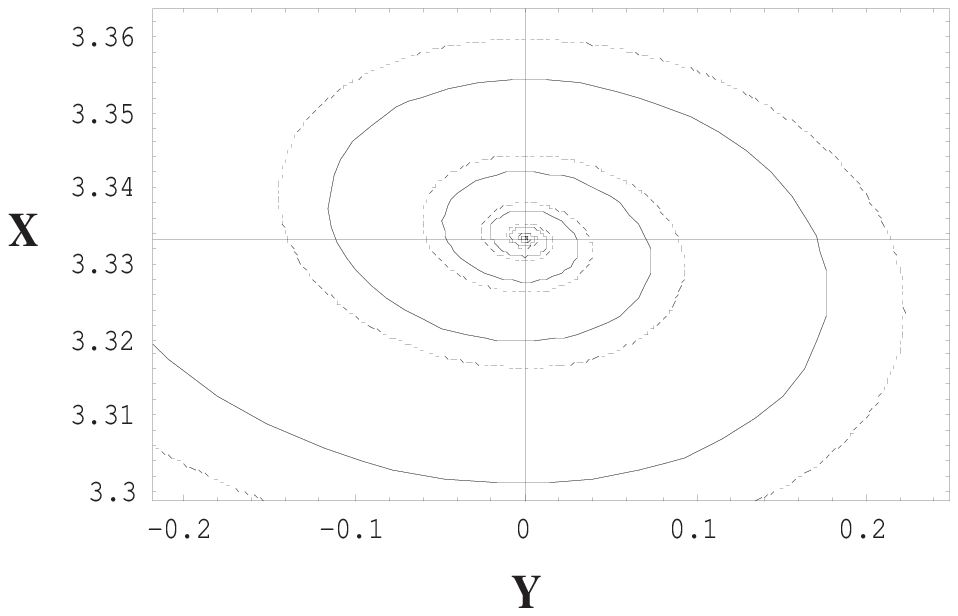}
\begin{center}{\small Fig.11 The phase portrait comparison for
quintessence with coupling term(real line) and without coupling term.
Both of them admit the same one attractor $X_c=3.33$.}
\end{center}
\end{center}
\vskip 0.3 in
\begin{flushleft}\textbf{{5. Conclusions}}\end{flushleft}
\par In this paper, we consider dilaton field as quintessence and phantom in dilatonic dark energy model.
We investigate the existence of a late time attractor in the the two
cases. The sufficient condition of the existence of attractor
solution is: the potential of the field has non-vanishing minimum
value in quintessence model, on the contrary, the potential should
have non-vanishing maximum value in phantom model. The physical
reason of the condition difference between quintessence and phantom
is clarified as follows: For a usual dynamic system(quintessence),
the system will be stable when the total energy of the system
arrives at the minimum value. So, for quintessence the sufficient
condition is that the potential of the field has non-vanishing
minimum value. However, for a phantom system whose kinetic energy is
negative, the system will be stable when the total energy of the
system arrives at the maximum value. So, for phantom the sufficient
condition is that the potential of the field has non-vanishing
maximum value.
\par The Mexican hat potential has a local minimum and a local maximum
which makes us consider quintessence and phantom in the same
potential. In quintessence model, the critical point is
$X_{c1}=\varepsilon$, and at this point the potential has a minimum
value $W_0$. In phantom model, we obtain that the critical point is
$X_{c2}=0$, and at this point the potential has a maximum value
$\frac{\mu\varepsilon^4}{4}+W_0$. We show mathematically that both
quintessence model and phantom model admits a late time attractor
solution corresponding to an equation of state $\omega=-1$(Fig.3 and
Fig.7) and a cosmic density parameter $\Omega_\sigma=1$(Fig.2 and
Fig.6), which are important features for a dark energy model that
can meet the current observations. Such evolution behaviors will
avoid the cosmic doomsday. Since
$T_{eq}\simeq5.64(\Omega_0h^2)eV\simeq2.843\times10^4K$,
$T_0\simeq2.7K$, $a_i=1$, the scale factor at the present epoch
$a_0$ would nearly be $1.053\times10^4$, then we know
$N_0=lna_0=9.262$. According to $N_0$, we obtain the current value
$\Omega_{\sigma,quintessence}\simeq0.707036$ and
$\Omega_{\sigma,phantom}\simeq0.703174$, which both meet the current
observations well.
\par Finally, we analyze the effect of the coupling term
$\frac{\alpha\eta_i e^{-\frac{1}{2}\alpha\sigma_0X}e^{-3N}}{4H}$ on
the evolution of $X$, $Y$ and attractor(Figs.9-11). We find that it
changes the evolutive process of $X$ and $Y$, but not the existence
of the attractor solution. These mathematical results show that,
because of the existence of this coupling between dilaton and
matter, the evolution of $X$ and $Y$ with respect to $N$ in
quintessence model will be faster than those of the other models
without coupling. According to the same analysis methods, we can
expect that the coupling term affects the evolution of $\omega$ and
$\Omega$ in the same way.
\par We can find that the state parameter ¦Ø of dilaton
as quintessence and phantom can both evolve to -1 from values
greater than -1 and from values smaller than -1, respectively. Can
it cross over -1? Or dilaton can be a candidate of quintom[31]?
Vikman has proved that the crossing of $\omega=-1$ is classically
forbidden in the single scalar field models[32]. However, in the
work of Singh [33], we can observe such a behavior of crossing of $w
= -1$ in semi-classical Loop Quantum Cosmology very naturally. If
one considers the quantum effects, such a crossing is possible for a
single scalar field.
\begin{flushleft}\textbf{Acknowledgements}\end{flushleft}
This work is partially supported by National Nature Science
Foundation of China under Grant No.10573012 and Shanghai Municipal
Science and Technology Commission No.04dz05905.

\begin{flushleft}{\noindent\bf References}
 \small{

\item {1.}{ A. G. Riess et al., \textit{Astrophys. J}\textbf{607}, 665(2004);
\\ \hspace{0.15 in}A. G. Riess, \textit{Astron. J}\textbf{116}, 1009(1998);
\\ \hspace{0.15 in}S. Perlmutter et al., \textit{Astrophys. J}\textbf{517}, 565(1999);
\\ \hspace{0.15 in}N. A Bahcall et al., \textit{Science}\textbf{284}, 1481(1999).}
\item {2.}{ C. L. Bennett et al., \textit{Astrophys. J. Lett}\textbf{148}, 1(2003).}
\item {3.}{ M. Tegmark et al., \textit{Phys. Rev. D}\textbf{69}, 103510(2004);
\\\hspace{0.15 in}M. Tegmark et al., \textit{Astrophys. J}\textbf{606}, 702(2004).}
\item {4.}{ J. S. Bagla, H. K. Jassal and T. Padmamabhan, \textit{Phys. Rev. D}\textbf{67}, 063504(2003).}
\item {5.}{ E. Elizalde, S. Nojiri and S. D. Odintsov, hep-th/0405034;
\\\hspace{0.15 in}S. Nojiri, S. D. Odintsov and M. Sasaki, hep-th0504052;
\\\hspace{0.15 in}S. Nojiri and S. D. Odintsov, hep-th/0601213.}
\item {6.}{ C. Wetterich \textit{Nucl. Phys. B}\textbf{302}, 668(1998);
\\\hspace{0.15 in}P. G. Ferreira and M. Joyce \textit{Phys. Rev. D}\textbf{58}, 023503(1998);
\\\hspace{0.15 in}J. Frieman, C. T. Hill, A. Stebbinsand and I.Waga, \textit{Phys. Rev. Lett}\textbf{75}, 2077(1995);
\\\hspace{0.15 in}P. Brax and J. Martin, \textit{Phys. Rev. D}\textbf{61}, 103502(2000)
\\\hspace{0.15 in}T. Barreiro, E. J. Copeland and N. J. Nunes, \textit{Phys. Rev. D}\textbf{61}, 127301(2000);
\\\hspace{0.15 in}I. Zlatev, L. Wang and P. J. Steinhardt \textit{Phys. Rev. Lett} \textbf{82}, 896(1999).}
\item {7.}{ T. Padmanabhan, and T. R. Choudhury, \textit{Phys. Rev. D}\textbf{66}, 081301(2002).}
\item {8.}{ A. Sen, \textit{JHEP} \textbf{0204}, 048(2002).}
\item {9.}{ C. Armendariz-Picon, T. Damour and V. Mukhanov, \textit{Phys. Lett. B}\textbf{458}, 209(1999).}
\item {10.}{A. Feinstein, \textit{Phys. Rev. D}\textbf{66}, 063511(2002);
\\\hspace{0.17 in}M. Fairbairn and M. H. Tytgat, arXiv:hep-th/0204070
\\\hspace{0.17 in}G. W. Gibbons,  arXiv:hep-th/0204008.}
\item {11.}{A. Frolov, L. Kofman and A. Starobinsky, \textit{Phys.Lett.B} \textbf{545}, 8(2002);
\\\hspace{0.17 in}L. Kofman and A. Linde, \textit{JHEP}\textbf{0207}, 004(2004).}
\item {12.}{C. Acatrinei and C. Sochichiu, \textit{Mod. Phys. Lett. A}\textbf{18}, 31(2003);
\\\hspace{0.17 in}S. H. Alexander, \textit{Phys. Rev. D}\textbf{65}, 0203507(2002).}
\item{13.}{T. Padmanabhan, \textit{Phys. Rev. D}\textbf{66}, 021301(2002).}
\item{14.}{A. Mazumadar, S. Panda and A. Perez-Lorenzana, \textit{Nucl. Phys. B}\textbf{614}, 101(2001);
\\\hspace{0.17 in}S. Sarangi and S. H. Tye,
\textit{Phys. Lett. B}\textbf{536}, 185(2002).}
\item{15.}{H. Q. Lu, \textit{Int. J. Mod. Phys. D}\textbf{14}, 355(2005);
\\\hspace{0.17 in}W. Fang, H. Q. Lu, Z. G. Huang and K. F. Zhang, \textit{Int. J. Mod. Phys. D}\textbf{15}, 199(2006).}
\item{16.}{X. Z. Li and J. G. Hao, \textit{Phys. Rev. D}\textbf{69}, 107303(2004).}
\item{17.}{T. Chiba, T. Okabe and M. Yamaguchi, \textit{Phys. Rev. D}\textbf{62}, 023511(2002).}
\item{18.}{P. Singh, M. Sami and N. Dadhich, arXiv:hep-th/0305110.}
\item{19.}{S. M. Carroll, M. Hoddman and M. Trodden, arXiv:astro-ph/0301273.}
\item{20.}{J. G. Hao and X. Z. Li, \textit{Phys. Rev. D}\textbf{68}, 083514(2003).}
\item{21.}{M. Sami, P. Singh, and S. Tsujikawa, arXiv:gr-qc/0605113}
\item{22.}{Michael Doran and J. J\"{a}chel, \textit{Phys. Rev. D}\textbf{66}, 043519(2002);
\\\hspace{0.17 in}N. J. Nunes and D. F. Mota, arXiv:astro-ph/0409481 and astro-ph/0504519;
\\\hspace{0.17 in}Ishwaree P. Neupane, arXiv:hep-th/0602097;
\\\hspace{0.17 in}Benedict M.N. Carter and Ishwaree P. Neupane, arXiv:hep-th/0512262 and hep-th/0510109;
\\\hspace{0.17 in}Federico Piazza and Shinji Tsujikawa, arXiv:hep-th/0405054;
\\\hspace{0.17 in}B. Gumjudpai, T. Naskar, M. Sami, and S. Tsujikawa, arXiv:hep-th/0502191.}
\item{23.}{L. Amendola, \textit{Phys. Rev. D}\textbf{62}, 043511(2000);
\\\hspace{0.17 in}E. Majerotto, D. Sapone and L. Amendola, arXiv:astro-ph/0410543.}
\item{24.}{H. Q. Lu, Z. G. Huang, W. Fang and K. F. Zhang, arXiv:hep-th/0409309.}
\item{25.}{H. Q. Lu and K. S. Cheng, \textit{Astrophysics and Space Science}\textbf{235}, 207(1996);
\\\hspace{0.17 in}Y. G. Gong, arXiv:gr-qc/9809015.}
\item{26.}{C. M. Will, arXiv:gr-qc/0103036.}
\item{27.}{T. Damour and K. Nordtvedt, \textit{Phys. Rev. Lett}\textbf{70}, 2217(1993).}
\item{28.}{B. Bertotti, L. Iess and P. Tortora, \textit{Nature}\textbf{425}, 374(2003).}
\item{29.}{Y .M .Cho, \textit{Phys. Rev.Lett}\textbf{68}, 3133(1992).}
\item{30.}{E. J. Copeland, M. Sami and S. Tsujikawa, arXiv:hep-th/0603057.}
\item{31.}{W. Hao, R. G. Cai and D. F. Zeng, \textit{Class.Quant.Grav}\textbf{22}, 3189(2005);
\\\hspace{0.17 in}Z. K. Guo, Y. S. Piao, X. M. Zhang, Y.Z. Zhang, \textit{Phys.Lett. B}\textbf{608}, 177(2005);
\\\hspace{0.17 in}B. Feng, arXiv:astro-ph/0602156.}
\item{32.}{A. Vikman, \textit{Phys. Rev. D}\textbf{71}, 023515(2005).}
\item{33.}{P. Singh, \textit{Class. Quant. Grav.}\textbf{22}, 4203(2005).}

}
\end{flushleft}
\end{document}